\documentclass[twocolumn,showkeys,superscriptaddress,nofootinbib,preprintnumbers,noshowpacs,tighte,showkeysnlines,notitlepage]{revtex4}
\usepackage[latin1]{inputenc}
\usepackage{amssymb}
\usepackage{color}
\usepackage{float}
\usepackage{dcolumn}
\usepackage{amsthm}
\usepackage[english]{babel}
\usepackage{hyperref,hypcap,ae,tikz}
\usepackage{graphicx}
\usepackage{bm}
\usepackage{amsmath}
\usepackage{mathrsfs}
\usepackage{mathtools}
\usepackage{amsfonts}
\usepackage[T1]{fontenc}
\usepackage{textcomp}
\usepackage{xcolor}
\usepackage{lmodern}
\usepackage[caption=false]{subfig}
\usepackage{booktabs}
\usepackage[section]{placeins}
\hypersetup{
	colorlinks=true,
	citecolor=blue,
    linkcolor=blue,
    urlcolor=blue,
	}

\begin{document}

\title{Ricci-Weyl curvature balance in viscous dissipative collapse: A covariant analysis of singularity censorship}

\author{Samarjit Chakraborty}
\email{samarjitxxx@gmail.com}

\author{Rituparno Goswami}
\email{Goswami@ukzn.ac.za}

\author{Sunil D. Maharaj}
\email{MAHARAJ@ukzn.ac.za}

\author{Gareth Amery}
\email{ameryg1@ukzn.ac.za}

\affiliation{Astrophysics Research Centre (ARC), College of Agriculture, Engineering and Science, University of KwaZulu-Natal (UKZN), Private Bag X54001, Durban 4000, South Africa.}
\affiliation{Discipline of Mathematics, School of Mathematics, Statistics and Computer Science, UKZN, Private Bag X54001, Durban 4000, South Africa.}

\begin{abstract}
We investigate the cosmic censorship conjecture in a spherically symmetric collapse with shear and bulk viscosity, heat flux, and pressure anisotropy, imposing physically reasonable energy conditions. Using the semi-tetrad covariant formalism, we derive the dynamics of the collapsing fluid, including a master equation for the evolution of the Weyl curvature, to examine the role of viscosity. The analysis of null geodesic geometry uncovers a novel curvature-balance mechanism between Ricci (matter) and Weyl (free gravitational field) curvature on the apparent horizon; this balance determines the causal nature of the horizon and thereby governs the visibility of the singularity. We then derive necessary and sufficient covariant conditions for the central singularity to be locally naked. Our findings support a weaker form of cosmic censorship and extend the covariant censorship analysis to realistic dissipative, viscous collapse.
\end{abstract}

\keywords{naked singularity; apparent horizon; gravitational collapse; semi-tetrad covariant formalism; cosmic censorship conjecture}

\maketitle

\section{Introduction}
In a general gravitational collapse the `final' end state is a black hole (BH) with an event horizon (EH) covering the central singularity. However, the EH is the global property of the spacetime and is only meaningful to an observer in infinitely distant future. Therefore, it was found (by Penrose\cite{M1}) that the concept of `trapped surfaces' is necessary, as they are locally defined and one need not wait for the entire evolution of the universe to study them. Essentially, due to the collapse of matter, the gravitational pull becomes so high that the local spacetime becomes `trapped'  with light rays always converging inside. The two dimensional surfaces that foliate such four dimensional trapped region are called `trapped surfaces'. Penrose showed in his famous \textit{singularity theorems} that under the null energy condition (which is least restrictive) without violating causality, if a `trapped surface' forms due to gravitational collapse, then the formation of singularity is inevitable. It is therefore obvious that if a trapped surface comes into existence before the formation of the central singularity, then the asymptotic structure of the spacetime remains safe with the singularity hidden behind the horizon. However, extreme situations may arise (mainly due to shear and density inhomogeneity) where the formation of the trapped surface is delayed and the singularity forms (from which light rays may or may not get trapped later) before the horizon. This can lead to serious issues in the global theorems of BH physics as they require the spacetime manifold to be \textit{future asymptotically predictable}. To prevent such `naked' singularity (to the future of the partial Cauchy surface), which is visible from the future null infinity, in 1969 Penrose proposed \cite{H1,H2} the famous \textit{cosmic censorship conjecture} (CCC). We can have two versions of this conjecture:

\vspace{0.1cm}
\textit{Strong CCC:} All physically reasonably spacetimes are globally hyperbolic, i.e., apart from possible initial singularities like Big Bang, no singularity is ever visible to any observer. Therefore, in a gravitational collapse  the singularity formed must remain (for all the time) in the trapped region and hence there cannot be any outgoing future directed non-spacelike geodesics from the vicinity of the singularity.

\vspace{0.1cm}
\textit{Weak CCC:} All singularities of a gravitational collapse are hidden within black holes and cannot be seen by a distant observer, admitting the possibility of the existence of local non-spacelike future directed geodesics. However, these rays must enter the trapped region in their future and fall back to the singularity to save the future asymptotic structure of the spacetime manifold (ruling out momentary global nakedness to save BH physics) \cite{R2}.

\vspace{0.1cm}
However, these are only conjectures, without any rigorous mathematical proof. Therefore, even if we assume that the central singularity is causally disconnected to the distant observer (in asymptotic future), a `local' visibility cannot be ruled out. Hence, any physically reasonable condition for a locally naked singularity would challenge the universal validity of strong CCC.

\subsection{Previous works}

Although no rigorous mathematical proof of CCC is available, many have tried to show (\cite{H3,H4} and the references therein) a number of counter examples where there are shell focusing naked singularities forming at the centre of the spherically symmetric dust, perfect fluid or radiation shell collapse. Extensive studies on various dynamical collapse models for a wide range of matter fields investigating the final outcome of gravitational collapse can be found in \cite{H5}. The generic conclusions which emerged from these studies suggest that the trapped surfaces may not develop early enough to always shield the interior high curvature region from an outside observer. Important insights also emerged from \cite{H6,H7}, where the authors showed that there exists a remarkable connection between spacetime shear and inhomogeneity of collapsing matter with the CCC. Essentially, the shear can distort the geometry of the trapped region in such a way that the central singularity can be locally naked. Also, it was shown in \cite{H19} that if a null geodesic originates from the central shell-focusing singularity, then there exist families of future-directed non-spacelike curves which can necessarily escape from there. Cosmic censorship in collapse was also investigated in \cite{H20,H21}. In recent years, studies were carried out on the trapped surfaces and the visibility of a central singularity in a covariant manner in \cite{L1,L2}. However, the study only deals with perfect collapsing fluids. Researchers have also examined the MOTS to be a possible location of Hawking radiation in \cite{Ellis}. Later, it was shown that thermodynamics of free gravitational field favours weak CCC in \cite{aqua}.

\subsection{This paper}

In light of the previous works, it becomes important to analyse the problem in a much more general settings and find out the necessary and sufficient conditions for the formation of a locally naked singularity during collapse. We aim to find a general model independent geometric condition and for that we adopt the semi-tetrad covariant approach. The novelty of this study relies upon the understanding of the curvature balance on the trapped horizon.
The Ricci-Weyl dichotomy reflects the fundamental split between the matter degrees of freedom (encoded in the Ricci tensor) and the pure gravitational degrees of freedom (encoded in the Weyl tensor). How this balance tips under extreme conditions reveals universal geometric criteria for locally naked singularity formation across different matter models. Therefore, the current work also validates and extends the results obtained in \cite{L1,L2}. To the best of our knowledge a comprehensive covariant study involving a physically rich collapsing fluid with dissipation, viscosity (bulk and shear) and pressure anisotropy in a shearing spherically symmetric spacetime is not currently available in literature. Therefore, we investigate the following question:

\vspace{0.5cm}
\textbf{Key question:} \textit{What is the role of Ricci-Weyl curvature balance in determining singularity visibility (locally) during viscous dissipative collapse, and what covariant conditions govern this balance?}

\vspace{0.1cm}
We impose physically reasonable energy conditions and study the slope of the tangent to the apparent horizon or the marginally outer trapped surface (MOTS) curve near the central singularity during the final epoch of the collapse to investigate the cause of locally naked singularity formation. Additionally, we explore the following problems in this work:

\begin{itemize}

\item \textit{How does the semi-tetrad covariant formalism describe viscous dissipative collapse dynamics? What equations govern Ricci-Weyl competition, and how does viscosity affect it?}
\item \textit{How does Ricci-Weyl balance influence MOTS evolution and causal structure?}
\item \textit{What covariant conditions make Weyl domination produce marginally naked singularities?}
\item \textit{What does the gravitational arrow of time reveal about curvature balance in censorship violation?}
\end{itemize}

The paper is organized as follows: In the next section we provide a brief description of the 1+1+2 semi-tetrad covariant decomposition for LRS II spacetimes. Then in section \ref{sec3} we set up the collapsing system in the covariant approach. Next, in section \ref{sec4}, we describe the governing dynamical equations of the collapsing fluid and derive the master equation for the Weyl scalar evolution. In section \ref{sec5} we use the null geodesics to analyse the trapped surface geometry. In section \ref{sec6} we examine the different types of MOTS and their relationship with curvature scalars. In section \ref{sec7} we finally derive the necessary and sufficient condition of the central singularity to be locally naked and examine some examples to strengthen our point. Lastly, in section \ref{sec8} we discuss various aspects of our results and possible consequences.
\section{Covariant description of LRS II spacetimes}
We employ the semi-tetrad 1+1+2 covariant description of LRS II spacetimes (the class which contains the spherically symmetric spacetimes) to study gravitational collapse. LRS II spacetimes are convenient to use as they are irrotational (with a continuous non-trivial isotropy group of spatial rotations at every point) and have a covariantly defined preferred spatial direction at each point. For detailed explanations please refer to \cite{H8,H9,H10,H11,H12,H13,H14}. 
We choose the timelike unit vector $u^a$ (usually defined along the fluid flow lines) and the orthogonal spacelike unit vector along the preferred spatial direction $e^a$, and decompose the spacetime as \cite{R18}
\begin{equation}\label{decomp}
g_{ab}=-u_au_b+e_ae_b+N_{ab},
\end{equation}
where $N_{ab}$ is the projection tensor on the spherical 2-shells and is defined as $ N_{ab}=h_{ab}-e_{a}e_{b} $. Here $ h_{ab} $ is the projection tensor (in 1+3 covariant decomposition) orthogonal to $ u^{a} $ and spans the $3$-space.

Therefore, the $1+3$ description $ \{u^{a}, h_{ab}\} $ gives rise to two kinds of derivatives: 
\begin{itemize}
\item \textbf{The dot derivative}: This is the covariant time derivative along the observers' worldlines . Therefore, for any tensor  $ S^{a...b}{}_{c...d}$, it is defined as 
$\dot{S}^{a...b}{}_{c...d}\, \equiv u^{e} \nabla_{e} {S}^{a...b}{}_{c...d} $.
\item \textbf{The \textit{D} derivative}: It is the orthogonally projected covariant spatial derivative onto the 3-space using the projection tensor $ h_{ab} $. Hence we define
$ D_{e}S^{a...b}{}_{c...d}{} \equiv h^a{}_f
h^p{}_c...h^b{}_g h^q{}_d h^r{}_e \nabla_{r} {S}^{f...g}{}_{p...q}$.
\end{itemize} 
In the $1+1+2$ formulation $\{u^{a}, e^{a}, N_{ab}\}$ we go a step further. The preferred spatial vector $e^a$ introduces two new derivatives by splitting the 3-space $(\text{therefore the}\, D\, \text{derivative})$, for any 3-tensor $ \psi_{a...b}{}^{c...d} $ :
\begin{itemize}
\item \textbf{The hat derivative}: This is the spatial derivative along the spacelike unit vector $e^a$ defined as $\hat{\psi}_{a...b}{}^{c...d} \equiv e^{f}D_{f}\psi_{a...b}{}^{c...d}$.
\item \textbf{The delta derivative}: It is the remaining projected spatial derivative on the 2-shell by the projection tensor $N_a^{~b}$ on all the free indices. Hence 
$\delta_f\psi_{a...b}{}^{c...d} \equiv N_{a}{}^{f}...N_{b}{}^gN_{h}{}^{c}...
N_{i}{}^{d}N_f{}^jD_j\psi_{f...g}{}^{i...j}$.
\end{itemize}

The covariant decomposition helps us to obtain a set of geometrical quantities for the chosen timelike congruence $ u^{a} $. These are the expansion scalar $\Theta$, acceleration 3-vector $\dot{u}^a$ and the shear 3-tensor $\sigma_{ab}$. The electric part of the Weyl tensor (responsible for tidal forces and inhomogeneity), can be similarly extracted as $E_{ab}= C_{acbd}u^cu^d$, whereas, the magnetic part of the Weyl tensor (due to  rotation or time varying  spacetime) $H_{ab}=C^*_{acbd}u^cu^d$, vanishes in spherical symmetry \cite{E7}, with $C^*_{abcd}$ as the right dual of the Weyl tensor. This timelike congruence $u^a$ also decomposes the energy momentum tensor of the matter field (EMT) into the energy density $\rho$, isotropic pressure $p$, heat flux 3-vector $q^a$ and the anisotropic stress 3-tensor $\pi_{ab}$. The only non-vanishing geometrical quantity related to the preferred spacelike congruence $ e^{a} $ is the volume expansion $\phi=\delta_ae^a$ due to spherical symmetry.

We can also obtain a set of covariant scalars from the above mentioned 3-vectors and 3-tensors as $\mathcal{A}=\dot{u}^ae_a$, $\Sigma=\sigma_{ab}e^ae^b$, $\mathcal{E}=E_{ab}e^ae^b$, $Q=q^ae_a$ and $\Pi=\pi_{ab}e^ae^b$. Therefore, the set of scalars that fully describe the spherically symmetric class of spacetimes are
\begin{equation}\label{set1}
\mathcal{D}\equiv\left\{\Theta, \mathcal{A}, \Sigma, \mathcal{E}, \phi, \rho, p, \Pi, Q\right\}.
\end{equation}
The aforementioned geometrical and thermodynamical scalars (from EMT), with their directional derivatives along $u^a$ (denoted by dot) and $e^a$ (denoted by hat) completely describe the Ricci and the Bianchi identities and therefore, can completely specify the dynamics of the spacetime. We can geometrically define the `gravitational mass' as the Misner-Sharp mass $ \mathcal{M}$, enclosed within the spherical $2$-shell at a given instant of time. In terms of the obtained covariant scalars it is expressed as \cite{E8,E9}
\begin{equation}\label{Mass}
\mathcal{M} = \frac{1}{2K^{\frac{3}{2}}}\left(\frac{1}{3}\rho - \mathcal{E} - \frac{1}{2}\Pi\right).
\end{equation}
Therefore, in vacuum, the `gravitational mass' is entirely due to the electric part of Weyl scalar $\cal E$. Here $K$ is the Gaussian curvature of the spherical 2-shells, which is related to the area radius of the shells $\mathcal{R}$ as the inverse square of $\mathcal{R}$. The Gaussian curvature $ K $ is related to the covariant scalars of Eq.(\ref{set1}) as  
\begin{equation}\label{gauss}
K \equiv\frac{1}{\mathcal{R}^2}=\frac{1}{3}\rho-\mathcal{E}-\frac{1}{2}\Pi +\frac{1}{4}\phi^2
-\left(\frac{1}{3}\Theta-\frac{1}{2}\Sigma\right)^2\, .
\end{equation}
The directional derivatives of $\mathcal{M}$ along $u^a$ and $e^a$ in terms of the covariant scalars are given by
\begin{eqnarray}
\label{MHat}
\hat{\mathcal{M}} &=& \frac{1}{4K^{\frac{3}{2}}}\left[\phi\rho - \left(\Sigma-\frac{2}{3}\Theta\right)Q\right],\\
\label{MDot}
\dot{\mathcal{M}} &=&  \frac{1}{4K^{\frac{3}{2}}}\left[\left(\Sigma - \frac{2}{3}\Theta\right)\left(p + \Pi\right) -\phi Q\right].
\end{eqnarray}

We now have all the necessary concepts and mathematical expressions to set up our collapsing model for further analysis.

\section{Nonadiabatic viscous collapse model}\label{sec3}

We represent the interior of a collapsing star using a general spherically symmetric metric in coordinates $(t, r, \theta, \phi)$ as
\begin{equation}
ds^2 = -A(t,r)^2 dt^2 + B(t,r)^2 dr^2 + C(t,r)^2 d\Omega^2
\end{equation}
where $A(t,r)$, $B(t,r)$, and $C(t,r)$ are arbitrary functions of time $t$ and radial coordinate $ r $. $d\Omega^2$ is the metric of the unit 2-sphere $d\Omega^2 = d\theta^2 + \sin^2\theta \, d\phi^2$. The Einstein field equations (in the natural units $\kappa=1$) are
\begin{equation}
G_{ab}\equiv R_{ab} - \frac{1}{2} R g_{ab}= T_{ab}.
\end{equation}
We can write (inside the stellar boundary $\mathcal{B}$) the covariant energy momentum tensor for a dissipative fluid with anisotropic pressure and viscosity in LRS II spacetimes as the following
\begin{equation}\label{EM1}
T_{ab}^{-}=\rho u_au_b + (p+\Pi)e_ae_b +2Qu_{(a}e_{b)}+\left(p-\frac12\Pi\right)N_{ab},
\end{equation}
where the effective radial pressure $(T_{ab}^{-}e^{a}e^{b})$ and tangential pressure $(T_{ab}^{-}N^{ab}/2)$ are given by the following 
\begin{equation}
P_{\parallel}\equiv p+\Pi=P_{r}-2\eta\Sigma-\zeta\Theta,\,\,\,\,P_{\perp}\equiv p-\frac{\Pi}{2}=P_{t}+\eta\Sigma-\zeta\Theta. 
\end{equation} 
Here $\eta$ and $\zeta$ denote the coefficient of shear and bulk viscosity respectively. $P_{r}$ and $ P_{t} $ are the radial and tangential components of the non-viscous pressure. The corresponding (covariant scalar) isotropic pressure $ p $ and anisotropic pressure $ \Pi $ becomes
\begin{align}
p=&\frac{1}{3}\left(P_{\parallel}+2P_{\perp}\right)=\frac{1}{3}(P_{r}+2P_{t})-\zeta\Theta,\nonumber\\
\Pi=&\frac{2}{3}\left(P_{\parallel}-P_{\perp}\right)=\frac{2}{3}\Delta P-2\eta\Sigma,
\end{align}
with $\Delta P\equiv(P_{r}-P_{t})$ as the non-viscous pressure anisotropy. Here $\rho$ is the energy density and $Q$ denotes the radial heat flux. In the $[u,e]$ plane (spherical 2-shells are trivial), the discriminant ($\Delta$) of the eigenvalue equations of the above EMT tensor (\ref{EM1}) takes the form
\begin{equation}\label{eigen1}
\Delta= (\rho+p+\Pi)^2-4Q^2\;.
\end{equation}
If the condition $\vert Q\vert < \frac12(\rho+p+\Pi)$ is satisfied, then there exists two distinct eigenvalues (one timelike and one spacelike eigenvector) on this plane, making it a Type I matter field(timelike matter field). However, when we have a vanishing discriminant
\begin{equation}\label{eigen2}
\vert Q\vert =\frac12(\rho+p+\Pi),
\end{equation}
it makes both the eigenvalues equal (with double degenerate null eigenvectors) and constrains the heat flux $Q$. Therefore, the resulting matter field becomes Type II (null).

We represent the four velocity of the fluid (comoving frame) by $u^{a}$ and take the radial unit vector as $e^{a}$ with the following 
\begin{align}
u^{a}u_{a}=-1, \;\; u^{a}e_{a}=0,\,\,\,e^{a}e_{a}=1, \nonumber\\
u^a= (A^{-1}, 0, 0, 0), e^a = (0, B^{-1}, 0, 0).
\label{4}
\end{align}
The corresponding kinematical scalars are 
\begin{align}
\mathcal{A}=\frac{\hat{A}}{A},\,\,\, \Theta =\dfrac{\dot{B}}{B}+2\dfrac{\dot{C}}{C},\,\,\,\Sigma=\frac{2}{3}\left(\frac{\dot{B}}{B}-\frac{\dot{C}}{C}\right),\,\,\,\phi=2\frac{\hat{C}}{C}.
\end{align}
From the field equations we can also write the matter variables in the following manner
\begin{align}
\rho-K =&\left(\frac{\dot{C}}{C}\right)^2-\left(\frac{\hat{C}}{C}\right)^2+2\frac{\dot{B}}{B}\frac{\dot{C}}{C}-2\frac{\hat{\hat{C}}}{C},\nonumber\\
\frac{1}{2}(\rho+3p)=&\left(\frac{\hat{\hat{A}}}{A}-\frac{\ddot{B}}{B}-2\frac{\ddot{C}}{C}+2\frac{\hat{A}}{A}\frac{\hat{C}}{C}\right),\nonumber\\
P_{\parallel}=&\left(\frac{\hat{C}}{C}\right)^2+2\frac{\hat{C}}{C}\frac{\hat{A}}{A}-\left(2\frac{\ddot{C}}{C}+\left(\frac{\dot{C}}{C}\right)^2\right),\nonumber\\
P_{\perp}=&\left(\frac{\hat{\hat{A}}}{A}+\frac{\hat{\hat{C}}}{C}+\frac{\hat{A}}{A}\frac{\hat{C}}{C}\right)-\left(\frac{\ddot{B}}{B}+\frac{\ddot{C}}{C}+\frac{\dot{B}}{B}\frac{\dot{C}}{C}\right),\nonumber\\
Q=&2\left(\frac{\hat{\dot{C}}}{C}-\frac{\dot{B}\hat{C}}{BC}\right).
\end{align}
For our chosen interior metric, the covariant hat and dot derivatives are defined as $ \dot{\mathcal{O}}\equiv 1/A (\partial \mathcal{O}/\partial t) $ and $ \hat{\mathcal{O}}\equiv 1/B(\partial \mathcal{O}/\partial r) $ for any scalar $ \mathcal{O} $. The Misner-Sharp mass for the interior spacetime is 
\begin{equation}
\mathcal{M}=\frac{C}{2}\left[\left({\dot{C}}^2-\hat{C}^2\right)+1\right].
\end{equation}

For outside the stellar boundary $\mathcal{B}$, we have the Vaidya spacetime (i.e., all outgoing radiation is massless) 
\begin{equation}
ds^2=-\left(1-\frac{2M(u)}{r}\right)du^2-2drdu+r^2(d\theta^2
+\sin^2\theta
d\phi^2) \label{1int},
\end{equation}
where $M(u)$ denotes the total gravitational mass inside the 2-shell and $u$ represents the retarded time. 
From the continuity of the first and second fundamental forms we have the following matching conditions
\begin{align}\label{junction}
\mathcal{M}(t,r)&\stackrel{\mathcal{B}}{=}M(u), \nonumber\\
M(u)_{,u}&\stackrel{\mathcal{B}}{=}-\frac{1}{2K}(1-2\mathcal{M}\sqrt{K})(P_{r}-2\eta\Sigma-\zeta\Theta).
\end{align}
Here, $\stackrel{\mathcal{B}}{=}$ means that both sides of the equation are evaluated on $\mathcal{B}$ and $ ,u $ denotes the partial derivative with respect to the retarded time coordinate. The matching of the full non-adiabatic sphere  (including viscosity) to the Vaidya spacetime is discussed in \cite{visc1, visc2}. We assume that physically reasonable energy conditions (i.e., the dominant energy condition (DEC)) are obeyed by the collapsing matter to preserve the causal structure of the spacetime manifold.
\begin{align}
&\text{\textbf{DEC:}}\,\,\,\rho\geq 0,\,\ \rho+P_{r}-2\eta\Sigma-\zeta\Theta \geq 2\mid Q \mid,\nonumber\\
&\,\ \rho+P_{t}+\eta\Sigma-\zeta\Theta \geq 0,
\, \rho \geq P_{r}-2\eta\Sigma-\zeta\Theta,\nonumber\\
& \, 2\rho(\rho+P_{r}-2\eta\Sigma-\zeta\Theta)\geq Q^2,\,\mid P_{t}+\eta\Sigma-\zeta\Theta \mid \leq \rho.
\end{align}

\section{Effect of viscosity on the thermodynamic variables}\label{sec4}

We now derive the equations governing the evolution and propagation of the covariant scalars that fully describe the LRS II spacetimes with shear and bulk viscosity. The viscosity contributions imply that we now have two new constraints on the system.
\begin{align}
\mathcal{C}_{1}\equiv \, & \Pi-\frac{2}{3}(P_{r}-P_{t})+2\eta\Sigma=0, \\
\mathcal{C}_{2}\equiv \, & p-\frac{1}{3}(P_{r}+2P_{t})+\zeta \Theta=0.
\end{align}
We demand that during the evolution these two constraints are obeyed, i.e., $\dot{\mathcal{C}_{1}}=0$ and $\dot{\mathcal{C}_{2}}=0  $.
The equations are derived using the Ricci identities for the vectors $u^a$ and $e^a$ and the doubly contracted Bianchi identities. Therefore the evolution equations for the covariant scalars become 
\begin{itemize}
\item\text{Evolution}:
\small \begin{align}
& 
\dot\phi = -\left(\Sigma-\dfrac{2}{3}\Theta\right)\left(\mathcal{A}-\dfrac{1}{2}\phi\right)
+Q, 
\label{phidot}
\\   
& 
 \dot\Sigma-\dfrac{2}{3}\dot\Theta  =
-\mathcal{A}\phi+2\left(\dfrac{1}{3}\Theta-\dfrac{1}{2}\Sigma\right)^2 
 \begin{multlined}[t] +\dfrac{1}{3}(\rho+2P_{r}+P_{t})\\
          -\zeta\Theta-\mathcal{E}-\eta\Sigma, 
 \end{multlined} \label{Sigthetadot}
\\  
& 
\dot{\mathcal{E}} -\dfrac{1}{3}\dot \rho+\dfrac{1}{3}\dot{(\Delta P)}-\dot{(\eta\Sigma)}= \begin{multlined}[t]
    \dfrac{1}{2}\left(3\mathcal{E}-P_{t}
    -\eta\Sigma-\rho
   +\zeta\Theta \right) \\
     \left(\Sigma-\dfrac{2}{3}\Theta\right) +\dfrac{1}{2}\phi Q.\end{multlined} \label{edot}
\end{align}
\end{itemize}
Similarly, if we demand in propagation that $ \hat{\mathcal{C}_{1}}=0 $ and $ \hat{\mathcal{C}_{2}}=0 $ are maintained, we obtain the following set of equations.
\begin{itemize}
\item\text{Propagation}:
\small\begin{align}
&\hat{\phi}  =  \begin{multlined}[t] -\dfrac{1}{2}\phi^2+\left(\dfrac{1}{3}\Theta+\Sigma\right)\left(\dfrac{2}{3}\Theta-\Sigma\right)-\dfrac{2}{3} \rho
    -\mathcal{E} \\
    -\dfrac{1}{3}\Delta P +\eta\Sigma,\end{multlined}\,\label{hatphinl}
\\  
&\hat{\Sigma}-\dfrac{2}{3}\hat{\Theta}=-\dfrac{3}{2}\phi\Sigma-Q,\label{Sigthetahat}
 \\  
&\hat{\mathcal{E}}-\dfrac{1}{3}\hat\rho+\dfrac{1}{3}\hat{(\Delta P)}-\hat{(\eta\Sigma)}= \begin{multlined}[t]
    -\dfrac{3}{2}\phi\left(\mathcal{E}+\dfrac{1}{3}\Delta P-\eta \Sigma\right) \\
    +\left(\dfrac{1}{2}\Sigma-\dfrac{1}{3}\Theta\right)Q. \end{multlined}
    \label{hateps}
\end{align}
\end{itemize}
The corresponding Raychaudhuri equation becomes
\small \begin{align}
&   \hat{\mathcal{A}}-\dot\Theta= \begin{multlined}[t] 
-\left(\mathcal{A}+\phi\right)\mathcal{A}+\dfrac{1}{3}\Theta^2
    +\dfrac{3}{2}\Sigma^2 +\dfrac{\rho}{2}
    +\dfrac{1}{2}\left(P_{r}+2P_{t}\right) \\
    -\dfrac{3}{2}\zeta\Theta.
 \end{multlined}   
    \end{align}
    \label{Raychaudhuri}
The propagation and evolution equation of heatflux are as follows
\begin{itemize}
\item\text{Propagation/evolution}:
\small \begin{align}
&    \dot\rho+\hat Q=\begin{multlined}[t] -\Theta\left(\rho+\dfrac{1}{3}(P_{r}+2P_{t})-\zeta\Theta\right)-\left(\phi+2\mathcal{A}\right)Q  \\
-\Sigma\Delta P+3\eta\Sigma^2,  
\end{multlined} 
    \label{Qhat}
    \end{align}  
 \begin{align}   
 \dot Q+\hat{P_{r}}-\hat{(\zeta\Theta)}-\hat{(2{\eta\Sigma})}
=&
-\phi\left(\Delta P-3\eta\Sigma\right) 
-\left(\dfrac{4}{3}\Theta+\Sigma\right) Q \nonumber\\
& -\left(\rho+P_{r}-\zeta\Theta-2\eta\Sigma\right)\mathcal{A}.
\end{align}
 \label{Qdot}
\end{itemize}
The aforementioned equations transparently demonstrate how the viscosity components influence the evolution and propagation of the kinematical and thermodynamical scalars. In fact, the shear and bulk viscosity terms influence each other via their evolution and propagation equations and hence they are not independent during collapse. 

Finally, we obtain the following master equation that the collapsing viscous fluid has to obey 
\small \begin{align}\label{master}
\dot{\mathcal{E}}-\frac{\dot{\rho}}{3}+\frac{1}{3}\dot{(\Delta P)}-\dot{\eta}\Sigma=\, &\eta\Big[\frac{2}{3}\hat{\mathcal{A}}+\frac{\mathcal{A}}{3}\left(2\mathcal{A}-\phi \right)-\Sigma\left(\frac{\Theta}{3}+\Sigma +\eta \right)\nonumber\\
&-\mathcal{E}+\frac{\Delta P}{3}\Big]
 +\frac{1}{2}\left(\Sigma - \frac{2}{3}\Theta\right)\big(3\mathcal{E}-P_{t} \nonumber\\
 &-\rho+\zeta\Theta \big)+\frac{1}{2}\phi Q.
\end{align}
The above equation can be obtained using the Raychaudhuri equation and the equation of shear evolution. From Eq.(\ref{Sigthetadot}) it is clear that the Weyl curvature is the source term for the shear evolution. This shear then deforms the apparent horizon and delays the trapping of the enclosed region. Hence the master equation (\ref{master}) has serious consequences on the formation of the trapped surfaces and the local `visibility' of the central shell-focusing singularity.
\section{Radial null geodesics and trapped surfaces in a viscous collapse}\label{sec5}
Let us consider light rays that are moving along the preferred spatial direction (radial null rays), characterised by the curves $x^{a}({\nu})$ with $\nu$ as the affine parameter of the null geodesic. The tangent vectors $ n^{a} $ to these curves are also null ( and propagate parallel to themselves) with the following properties 
\begin{equation}
n^{b}\nabla_{b}n^{a}=\frac{\delta n^{a}}{\delta \nu}=0,\,\,\, n^{a}n_{a}=0.
\end{equation}
As the light rays move along the preferred spatial direction, the components of these null curves on the spherical 2-shells vanish. If $ e^{a}n_{a}>0 $ then the geodesics are locally outgoing and $e^{a}n_{a}<0  $ means a locally ingoing geodesic. In spherical symmetry the equation of the tangents to the outgoing (ingoing) null geodesics along the preferred spatial direction are then
\begin{equation}
k_{a}=\frac{E}{\sqrt{2}}(u_{a}+e_{a}),\,\,\, l^{a}=\frac{1}{\sqrt{2}E}(u^{a}-e^{a}),\,\,\, k_{a}l^{a}=-1,
\end{equation}
where $E$ is the energy of the light ray. Using the full covariant derivative of $u_a$ and $e_a$ we can obtain the volume expansion scalar of the outgoing null rays as \cite{H15,H16}
\begin{equation}
\tilde{\Theta}_{out}=\frac{E}{\sqrt{2}}N^{ab}\nabla_{a}(u_{a}+e_{b})=\frac{E}{\sqrt{2}}\left(\frac{2}{3}\Theta-\Sigma+\phi\right).
\end{equation}
Similarly, for the ingoing null rays the expansion is
\begin{equation}
\tilde{\Theta}_{in}=\frac{1}{\sqrt{2}E}\left(\frac{2}{3}\Theta-\Sigma-\phi\right).
\end{equation} 
\subsection{Geometry of trapped surfaces and MOTS}
Let us now consider a spherical emitter of light, surrounding a massive object. If sufficiently large amounts of matter is present within that emitting sphere, the volume expansion of the outgoing null congruence $(\tilde{\Theta}_{out})$ orthogonal to the sphere becomes negative. Therefore, both the outgoing and ingoing wavefronts collapse towards the central region. This emitting sphere is then called a closed trapped 2-surface. The collection of all such closed trapped 2-surfaces (in four dimensional manifold) is called a 4 dimensional trapped region. The 3 dimensional boundary of the trapped region is called a marginally outer trapped surface (MOTS), where the expansion of the outgoing null rays vanish with converging ingoing null rays. Due to spherical symmetry it is sufficient to study the one dimensional MOTS curve in the local $\{u^{a},e^{a}\}$ plane to determine its local properties. We now obtain the curve that describes the MOTS by $\tilde{\Theta}_{out}=0 $ as the following\cite{H18}:
\begin{equation}
\Psi\equiv \left(\frac{2}{3}\Theta-\Sigma+\phi\right)=0,\Rightarrow 
\dot{K}+\hat{K} \bigg\vert_{\Psi=0}=0,\,\,\,\dot{K}=-\hat{K}=\phi K.
\end{equation}
Therefore, on the MOTS, we get $ \tilde{\Theta}_{in}=-\sqrt{2}\phi/E $. The Gaussian curvature of the spherical shells is assumed to be decreasing along the integral curves of the preferred spatial direction and, therefore, from the relation $ \hat{K}=-\phi K $ a positive spatial volume expansion of the preferred spatial direction ($\phi > 0$) within the collapsing star is the only option. This is a sufficient condition to avoid any shell crossing singularities during the collapse. Hence on the MOTS (and within) we have $\tilde{\Theta}_{in}<0$. Consequently, an interesting geometrical property (in spherical symmetry) on the MOTS is the vanishing of the 4-gradient of the local Gaussian curvature of any 2-shell. Therefore, the Misner-Sharp mass enclosed by the 2-shell becomes
\begin{equation}
\mathcal{M}=\frac{1}{2\sqrt{K}}\left[1-\frac{1}{4K^3}\nabla_{a}K \nabla^{a}K\right] \bigg\vert_{\Psi=0},\,\,\, \Rightarrow\,\,\, 2\mathcal{M}\sqrt{K}=1.
\end{equation}

\vspace{0.15cm}
{\raggedright
\textbf{Definition:} \textit{A smooth, three-dimensional submanifold $\mathscr{H}$ in a spacetime $(\mathscr{M},g)$ is said to be a marginally outer trapped surface (MOTS) if it is foliated by a preferred family of 2-spheres such that, on each leaf $\mathscr{S}$, the expansion $\tilde{\Theta}_{out}$ of the outgoing null normal $ k^{a} $ vanishes and the expansion $ \tilde{\Theta}_{in} $ of the ingoing null normal $ l^{a} $ is strictly negative. In other words, the necessary and sufficient condition for any sub-manifold $\mathscr{H}$ to be the MOTS is that the gravitational mass enclosed by each leaf $\mathscr{S}$ is half the area radius of the leaf.}
}

\vspace{0.15cm}
This implies that, on the MOTS curve for any 2-shell, the enclosed gravitational mass is twice its area radius, which also corresponds to the black hole horizon radius. We can employ this feature to identify MOTS in any spacetime. For a detailed analysis of the aforementioned properties of MOTS see \cite{L1,L2}.
In spherical symmetry (for any shell) we have the following geometric relation which shows a balance between the kinematical and thermodynamic variables. This balance is controlled by the ratio of the gravitational mass and area radius of that shell, known as the `compactness' $\mathcal{C}\equiv \mathcal{M}/\mathcal{R}$ \cite{virial}.

\begin{equation}
\phi^2-\left(\Sigma-\frac{2}{3}\Theta\right)^2=\frac{2(1-2\mathcal{C})}{\mathcal{C}}\left[\frac{\rho}{3}-\mathcal{E}-\frac{\Pi}{2}\right],\,\,\,\,\mathcal{C}\equiv \mathcal{M}\sqrt{K}.
\end{equation}
In terms of the expansion scalars of the outgoing and ingoing null vectors we can express the above equation as
\begin{equation}
\frac{1}{2}\tilde{\Theta}_{in}\tilde{\Theta}_{out}=\left(2\mathcal{M}\sqrt{K}-1 \right)K.
\end{equation}
Therefore, (outside the trapped surface) when $\tilde{\Theta}_{in}<0$ and $\tilde{\Theta}_{out}>0 $, it automatically implies $\mathcal{R}>2\mathcal{M}$. On the marginal trapped surface $\tilde{\Theta}_{out}=0$, i.e., $  \mathcal{R}=2\mathcal{M}$, and inside the trapped surface ($\tilde{\Theta}_{in}<0$ and $ \tilde{\Theta}_{out}<0$) $ \mathcal{R}<2\mathcal{M} $. On the MOTS $ (\Psi=0) $ the right hand side of the above relation vanishes.

\vspace{0.5cm}
We also compute the covariant time and spatial derivatives of the gravitational mass on the MOTS:
\begin{align}
\dot{\mathcal{M}}\vert_{\Psi=0}=&\frac{\phi}{4K^{3/2}}(P_{r}-2\eta\Sigma-\zeta\Theta-Q),\nonumber\\
\hat{\mathcal{M}}\vert_{\Psi=0}=&\frac{\phi}{4K^{3/2}}(\rho-Q).
\end{align}
The above results have important implications. The first equation essentially tells us that if the effective radial pressure is equal to the heat flux on the MOTS, then the gravitational mass has no time dependence. In the case where dissipation dominates over the radial pressure on the MOTS, we have $ \dot{\mathcal{M}}<0 $. The second equation implies that 
under physically reasonable energy conditions (where $ Q< \rho $)
$ \hat{\mathcal{M}}>0 $ on the MOTS and equality of heat flux and energy density makes the gravitational mass only time dependent.  
The complete covariant derivative of the Misner-Sharp mass gives us the following
\small \begin{align}\label{massMOTS}
\nabla_{a}\mathcal{M}\nabla^{a}\mathcal{M}\vert_{\Psi=0}=\frac{\phi^2}{16K^3}(\rho+P_{r}-2\eta\Sigma-\zeta\Theta-2Q)\nonumber\\
(\rho-P_{r}+2\eta\Sigma+\zeta\Theta).
\end{align}
Therefore, on the MOTS, the 4-gradient of the enclosed gravitational mass becomes null for Type II matter fields ($-\hat{\mathcal{M}}=\dot{\mathcal{M}}$) or when the effective radial pressure is equal to the energy density (could be Type I or Type II with $\hat{\mathcal{M}}=\dot{\mathcal{M}}$). The trivial $ \phi=0 $ condition corresponds to the Schwarzschild black hole horizon. In general, the right hand side is always positive under physically reasonable energy conditions for Type I matter fields.

We can always write the tangent vector to the $ \Psi=0 $ MOTS curve in the $\{u^{a},e^{a}\}$ plane as $ \Psi^{a}=\alpha u^{a}+\beta e^{a}$. We also know the normal vector can be expressed as $\nabla_{a}\Psi=-\dot{\Psi}u_{a}+\hat{\Psi}e_{a}$. Therefore, we obtain the relation $ \Psi^{a}\nabla_{a}\Psi=\alpha \dot{\Psi}+\beta \hat{\Psi}=0 $. Therefore, in general, the norm of the tangent on the MOTS is 
\small \begin{align}\label{tangNorm}
\Psi^{a}\Psi_{a}=\,\beta^2\left(1-\frac{\alpha^2}{\beta^2}\right)= &-\frac{\beta^2}{3(P_{r}-2\eta\Sigma-\zeta\Theta+K-Q)^2}(R+6\mathcal{E}) \nonumber\\
&(\rho+P_{r}-2\eta\Sigma-\zeta\Theta-2Q),
\end{align}
where $ R=\rho-(P_{r}+2P_{t})+3\zeta\Theta=-T $ is the Ricci scalar, and $T$ is the trace of the matter EMT. The above equation transparently shows the interplay between the Ricci scalar and the Weyl scalar (timelike/spacelike MOTS accordingly) during collapse and how they come to a specific balance when a null MOTS forms. It is clear that if the collapsing matter remains Type I then due to the energy conditions the outward heatflux has to be $ Q<(\rho+P_{r}-2\eta\Sigma-\zeta\Theta)/2 $. However, in the Type II limit the heatflux approaches the value $Q=(\rho+P_{r}-2\eta\Sigma-\zeta\Theta)/2$. Therefore, under physically reasonable conditions the last term in the parenthesis can never be negative.

\vspace{0.3cm}
{\raggedright
\textbf{Proposition 1:} \textit{Consider the continued collapse of a general spherically symmetric viscous fluid that obeys the physically reasonable energy conditions from a regular initial epoch. If the following conditions are satisfied: \\
1. The spacetime is free of shell crossing singularities, i.e., $ \phi>0 $,\\
2. Closed trapped surfaces exist,\\
then the only local quantity that determines the causal nature of the MOTS is $R+6\mathcal{E}$ or $ -T+6\mathcal{E}$. }
}

\section{How would the MOTS evolve in presence of matter?}\label{sec6}

In order to understand the notion of future outgoing (ingoing) MOTS we examine the ratio $\alpha/\beta$ (slope of the tangent to the MOTS curve). If $\alpha/\beta > 0$ then the MOTS is said to be future outgoing and if it is less than zero then the MOTS is future ingoing. The general expression of this ratio is
\begin{equation}
\frac{\alpha}{\beta}=-\frac{\hat{\Psi}}{\dot{\Psi}}=\dfrac{2\rho/3+\Delta P/3-\eta\Sigma+\mathcal{E}-Q}{-\rho/3-(2P_{r}+P_{t})/3+\zeta\Theta+\eta\Sigma+\mathcal{E}+Q}.
\end{equation}
It is physically reasonable to think that the heat flux $Q$ is relatively weak compared to other thermodynamic quantities (and obey the energy conditions). We express the Weyl scalar in terms of the Gaussian curvature $K$ of the 2-shell, energy density and anisotropy to write the above relation on the MOTS curve as 
\begin{equation}\label{ratio1}
\frac{\alpha}{\beta}=\frac{-\rho+K+Q}{P_{r}-2\eta\Sigma-\zeta\Theta+K-Q}\leq 1.\,\,\, \text{(DEC)}.
\end{equation}

\vspace{0.1cm}
{\raggedright
\textbf{Corollary 1:}\textit{ Physically reasonable energy conditions constrain the ratio $ \alpha/\beta $ to less than unity for Type I matter, and equality holds for Type II field. }}

\vspace{0.1cm}
Consequently, we can also write
\begin{equation}
\frac{\alpha}{\beta}=\dfrac{\phi-2\hat{\mathcal{M}}/\mathcal{M}}{\phi+2\dot{\mathcal{M}}/\mathcal{M}}\leq 1. 
\end{equation}
Due to the energy conditions we get an interesting constraint on the enclosed gravitational mass. If $ \dot{\mathcal{M}}\geq 0 $ (accretion), then we must have $\hat{\mathcal{M}}+\dot{\mathcal{M}}\geq 0  $. In case of dissipation if $ \dot{\mathcal{M}}<-\phi/4\sqrt{K} $ then it automatically implies $ \hat{\mathcal{M}}+\dot{\mathcal{M}}\leq 0 $. In the above expression when equality holds (in energy condition), we have an outgoing null MOTS, since $ \alpha/\beta=1 $ (see the \textbf{Corollary 2 and 3}) with $\hat{\mathcal{M}}+\dot{\mathcal{M}}=0$. It is interesting to notice that for future ingoing null MOTS the ratio $ \alpha/\beta=-1 $ corresponds to $ \phi=2\sqrt{K}(\hat{\mathcal{M}}+\dot{\mathcal{M}})$. As both the temporal and spatial covariant derivatives of $ \mathcal{M} $ vanish, the MOTS approaches an end static state of a collapse \cite{H17} with $ \nabla_{a}\mathcal{M}=0 $, like an eternal black hole, as shown in the following

\begin{equation}
\text{on}\,\, \Psi=0:\,\,\,\,\,\,\frac{\alpha}{\beta}\rightarrow 1,\,\,\,\Rightarrow \hat{\mathcal{M}}\rightarrow 0,\,\,\, \text{and}\,\,\,\, \dot{\mathcal{M}}\rightarrow 0.
\end{equation}

Therefore, for an initial null marginally outer trapped surface \( S_{\mathrm{MOTS}} \) \cite{Ellis} to remain null in the presence of matter, the matter field must be either Type II or a Type I field satisfying \( 2\mathcal{E} = (P_{r} + 2P_{t})/3 - \zeta\Theta - \rho/3 \). This leads to the following proposition.

\vspace{0.2cm}
{\raggedright\textbf{Proposition 2:}\textit{ Consider the continued collapse of a general spherically symmetric viscous fluid that obeys the physically reasonable energy conditions from a regular initial epoch. If the following conditions are satisfied: \\
1. The spacetime is free of shell crossing singularities, i.e., $ \phi>0 $,\\
2. Closed trapped surfaces exist,\\
then an initial null MOTS can only remain null, if the infalling matter satisfies $R+6\mathcal{E}=0$, i.e., $-T+6\mathcal{E}=0$  or reaches the Type II condition with an outward Type II heatflux 
\begin{equation}
Q=\frac{\rho+P_{r}-2\eta\Sigma-\zeta\Theta}{2},
\end{equation}
and if matter remains Type I with $R+6\mathcal{E}\neq 0$ , i.e., $6\mathcal{E}\neq T$ then the MOTS will cease to be null and change to spacelike or timelike. }}

\vspace{0.1cm}
This is similar to the findings obtained in \cite{DG}, where the authors showed the condition of preserving the degeneracy of the event horizon and the null MOTS requires a limiting Vaidya radiation, which helps the eventual Hawking radiation.

\subsection{Null MOTS}

For a Type I matter field the condition for a null MOTS simply reduces to
\begin{equation}
\Psi^{a}\Psi_{a}=0 \Rightarrow R+6\mathcal{E}=0,\,\,\Rightarrow\,\,\, 6\mathcal{E}=T.
\end{equation}
From the above condition we can obtain the equation for a null MOTS for Type I matter fields as

\begin{equation}
\rho-\frac{4\mathcal{M}}{\mathcal{R}^3}-P_{r}+\zeta\Theta+2\eta\Sigma \bigg\vert_{\Psi=0}=0,
\end{equation}
or equivalently,
\begin{equation}
\rho-\frac{4\mathcal{M}}{\mathcal{R}^3}\stackrel{\Psi=0}{=}P_{\parallel}.
\end{equation}
For a null MOTS the required condition is $\alpha^2/\beta^2=1 $. Therefore, we have two possibilities, $\alpha/\beta=\pm 1$, i.e., $\dot{\Psi}=\mp \hat{\Psi}$. In terms of the covariant scalars we get the following 
\begin{align}
&\hat{\Psi}=\dot{\Psi},\,\,\,\Rightarrow\,\,\,\,R+6\mathcal{E}=0,\nonumber\\ 
& -\hat{\Psi}=\dot{\Psi},\,\,\,\Rightarrow\,\,\,\,\rho+P_{r}-2\eta\Sigma-\zeta\Theta=2Q.
\end{align}
{\raggedright
\textbf{Corollary 2:} A \textit{future outgoing null MOTS is possible only with a Type II matter field.}}

The future ingoing null MOTS corresponds to the condition $ R+6\mathcal{E}=0 $ as $ \alpha/\beta $ is negative. We also observe that, in Eq.(\ref{tangNorm}), if the weak energy condition is satisfied avoiding the equality, i.e., the matter is of Type I, then $ 2\mathcal{E}=(P_{r}+2P_{t})/3-\zeta\Theta-\rho/3 $ implies that the MOTS is null.

\vspace{0.2cm}
{\raggedright
\textbf{Corollary 3:} \textit{Under physically reasonable energy conditions for a Type I matter field with $ R+6\mathcal{E}=0 $, i.e., $-T+6\mathcal{E}=0$, the null MOTS is always future ingoing.}}

\vspace{0.2cm}
\subsection{Non-null MOTS}
{\raggedright\textbf{Proposition 3:}\textit{ Consider the continued collapse of a general spherically symmetric viscous fluid that obeys the physically reasonable energy conditions from a regular initial epoch. If the following conditions are satisfied: \\
1. The spacetime is free of shell crossing singularities, i.e., $ \phi>0 $,\\
2. Closed trapped surfaces exist,\\
then only ingoing timelike MOTS $(\alpha/\beta<-1)$ are possible. Whereas, spacelike MOTS can be both ingoing or outgoing $(-1<\alpha/\beta<1)$ depending upon the sign of $\alpha/\beta$.}}

A timelike MOTS is possible when $ \alpha^2 /\beta^2>1$. Therefore, if the ratio ($\alpha/\beta>0$) is positive, it implies $\alpha/\beta>1$. However, greater than unity (future outgoing timelike MOTS) values violate the energy conditions (see Eq.(\ref{ratio1})). Thus, a timelike MOTS can only be a future ingoing MOTS for physically reasonable energy conditions with  $ \alpha/\beta<-1 $ and satisfy the following relation
\begin{equation}
K<\frac{1}{2}(\rho-P_{r}+2\eta\Sigma+\zeta\Theta),\,\,\,\,\Rightarrow R+6\mathcal{E}>0,\,\,\,\Rightarrow 6\mathcal{E}>T.
\end{equation}
In general such timelike trapped surfaces usually correspond to the inner MOTS (IMOTS) which are future ingoing. For more details see \cite{Ellis}.

For spacelike MOTS the allowed range is $ -1<\alpha/\beta<1 $, meaning it can be outgoing or ingoing depending on the sign. It is interesting to see that $ \alpha/\beta<1 $ automatically implies that the matter should be of Type I as $  2Q<\rho+P_{r}-2\eta\Sigma-\zeta\Theta $. Therefore, a spacelike MOTS with a Type I matter field (with $ \alpha/\beta>-1 $ condition) obeys the following:
\begin{equation}
K>\frac{1}{2}(\rho-P_{r}+2\eta\Sigma+\zeta\Theta),\,\,\,\,\Rightarrow R+6\mathcal{E}<0,\,\,\,\Rightarrow 6\mathcal{E}<T.
\end{equation}
It is clear that the bounds of the Gaussian curvature are consistent with \textbf{Proposition 1 and 2}. 

\vspace{0.2cm}
{\raggedright
\textbf{Corollary 4:} \textit{Under physically reasonable energy conditions during a viscous dissipative collapse, the Weyl scalar on the future ingoing MOTS is bounded, whereas for the future outgoing MOTS it is not.}}

In general, we can make some observations about the Weyl scalar on the MOTS. If $ \alpha/\beta>0 $, the future outgoing MOTS corresponds to either one the two following conditions $ K>\rho-Q $ or $ K<Q-P_{r}+2\eta\Sigma+\zeta\Theta $. In terms of Weyl curvature this relation reads
\begin{equation}\label{weyl_bound}
\mathcal{E}<Q-2\rho/3-\Delta P/3+\eta\Sigma,\,\,\text{or}\,\,\,\mathcal{E}>\dfrac{1}{3}(2P_{r}+P_{t})-\zeta\Theta-\eta\Sigma+\frac{\rho}{3}-Q,
\end{equation}
where the second condition will play a decisive role in the end state of the collapse regarding the visibility of the central singularity. And when $ \alpha/\beta<0 $ (future ingoing MOTS), it corresponds to $ Q-P_{r}+2\eta\Sigma+\zeta\Theta<K<\rho-Q $ or equivalently
 \begin{equation}
Q-\dfrac{2\rho}{3}-\dfrac{\Delta P}{3}+\eta\Sigma<\mathcal{E}<\dfrac{(2P_{r}+P_{t})}{3}-\zeta\Theta-\eta\Sigma+\dfrac{\rho}{3}-Q.
 \end{equation}
Hence, \textbf{Corollary 4} is proved.

\section{Singularity}\label{sec7}
The spacetime point on which the central shell becomes singular must lie on the MOTS. If there exists an open set $ \mathscr{U} $ in $(\mathscr{M},g)$ that contains the outgoing null rays with their past end points arbitrarily close to a spacetime singularity and the null rays do not hit the MOTS in that open set, then the singularity is said to be locally naked in that open set (in the neighbourhood of the central shell-focusing singularity) \cite{H19}. Previously in \cite{L1,L2}, it was proved that the necessary and sufficient condition for the existence of an open set  in which the central singularity is locally naked, is that the MOTS at the central singularity should be future outgoing and non-spacelike (i.e., $6\mathcal{E}\geq T$) with $ \alpha/\beta\geq 1 $.

\vspace{0.2cm}
{\raggedright\textbf{Proposition 4:}\textit{ Consider the continued collapse of a general spherically symmetric viscous fluid that obeys the physically reasonable energy conditions from a regular initial epoch. If the following conditions are satisfied: \\
1. The spacetime is free of shell crossing singularities, i.e., $ \phi>0 $,\\
2. Closed trapped surfaces exist,\\
3. An open set exists that contains the outgoing null rays with their past end points arbitrarily close to the spacetime singularity and the null rays do not hit the MOTS in that open set,\\
then the central shell focusing singularity can only be marginally naked (locally). }}

From the last argument it is clear that to have a locally naked singularity we need the MOTS to be either timelike or null and the MOTS needs to be future outgoing at the central singularity. However, our \textbf{Proposition 3} already showed that future outgoing timelike MOTS 
($\alpha/\beta>1$) violates energy conditions. Therefore, the only viable possibility is that the central singularity is marginally naked with null future outgoing MOTS.

\vspace{0.2cm}
{\raggedright\textbf{Corollary 5:}\textit{ In a physically reasonable gravitational collapse Weak CCC is preserved with a  marginally locally naked central singularity with a future outgoing null MOTS. }}

Now, we recall our \textbf{Corollary 2}, where we proved that future outgoing null MOTS is possible only with a Type II matter field. Therefore, the central singularity can be locally naked if at the local MOTS (outgoing null) matter becomes Type II (respecting physically reasonable energy conditions). However, we can think of a scenario where due to extreme collapse conditions near the singularity standard matter is reaching the limit of Type II (classically) and the quantum processes take over.

\subsubsection{Marginally locally naked singularity}

Let us consider a scenario in general where the central shell-focusing singularity is marginally naked $ \alpha/\beta=1 $. Therefore, we have the following relation.
\begin{align}
& \Bigg[\frac{\mathcal{E}}{\rho+P_{r}-2\eta\Sigma-\zeta\Theta-2Q} \nonumber\\
& -\frac{\rho+(2P_{r}+P_{t})-3\zeta\Theta-3\eta\Sigma-3Q}{3(\rho+P_{r}-2\eta\Sigma-\zeta\Theta-2Q)}\Bigg]^{-1}=0.
\end{align}
It is clear that for Type II matter the above relation is satisfied trivially. However, under physically reasonable energy conditions, the collapsing matter field can reach extreme conditions where all the thermodynamic scalars diverge, along with the non-trivial quantum gravity processes. Interestingly, the ratio 
$ {\rho+(2P_{r}+P_{t})-3\zeta\Theta-3\eta\Sigma-3Q}/{3(\rho+P_{r}-2\eta\Sigma-\zeta\Theta-2Q)}$ remains finite in this limit due to the energy conditions (\textbf{DEC}). Therefore, under such extreme conditions the only possible solution would be a diverging $\vert\mathcal{E}\vert/\rho+P_{r}-2\eta\Sigma-\zeta\Theta-2Q $. Hence, our next proposition follows.

\vspace{0.2cm}
{\raggedright\textbf{Proposition 5:}\textit{ Consider the continued collapse of a general spherically symmetric viscous fluid that obeys the physically reasonable energy conditions from a regular initial epoch. If the following conditions are satisfied: \\
1. The spacetime is free of shell crossing singularities, i.e., $ \phi>0 $,\\
2. Closed trapped surfaces exist,\\
3. Central shell-focusing singularity is marginally naked,\\
then in the vicinity of the singularity on the non-spacelike outgoing MOTS the limiting value of $\dfrac{\vert\mathcal{E}\vert}{\rho+P_{r}-2\eta\Sigma-\zeta\Theta-2Q} $ diverges, implying that\\
\vspace{0.1cm}
1. Either the Weyl scalar $ \mathcal{E} $ completely dominates over the matter variables,\\
2. Or the matter becomes Type II,\\
3. Or both the above conditions satisfy simultaneously.}}
\vspace{0.2cm}

This seems to be the only viable condition for the existence of a marginally locally naked singularity. Therefore, under extreme collapse conditions near the central shell-focusing singularity the Weyl curvature plays a crucial role. If by any mechanism the Weyl curvature completely dominates over the matter curvature terms on the MOTS in the neighbourhood of the singularity (tidal effects deforming the geometry of the trapped surface and delaying its formation) we will have a marginally locally naked singularity.
\subsubsection{Perfect fluid with viscosity}

We set $ Q=\Delta P=0 $ to make the system non-dissipative and the non-viscous part of the anisotropic pressure vanish. We want to study such a system to understand the specific role of viscosity terms during collapse. Interestingly, we can determine the coefficient of shear viscosity completely in terms of our known covariant quantities

\begin{equation}
\eta=\dfrac{1}{3}\dfrac{\hat{P_{\parallel}}+(\rho+P_{\parallel})\mathcal{A}}{\Sigma \phi}=-\dfrac{P_{\parallel}-P_{\perp}}{3\Sigma}\,.
\end{equation}
Consequently, we get the following relation, which is nothing but the last propagation/evolution equation discussed in the previous section.
\begin{equation}
\hat{P_{\parallel}}+(\rho+P_{\parallel})\mathcal{A}+\phi P_{\parallel}=\phi P_{\perp}.
\end{equation}
In this case the ratio of $\alpha$ and $\beta$ becomes
\begin{equation}
\dfrac{\alpha}{\beta}=\dfrac{2\rho/3-\eta\Sigma+\mathcal{E}}{-\rho/3-P_{i}+\zeta \Theta+\mathcal{E}+\eta\Sigma},\,\,\,P_{i}=P_{r}=P_{t}.
\end{equation}
If we consider the future outgoing null MOTS with $\alpha/\beta=1$, then along the MOTS curve the Weyl scalar dominates over the inertial mass to make the ratio $ \vert{\mathcal{E}}\vert/\vert{\rho+P_{\parallel}}\vert $ diverge, making the central singularity marginally locally naked.

If we now additionally impose the conformally flat condition $ \mathcal{E}=0 $, we obtain the norm of tangent to the MOTS as

\begin{equation}
\Psi^{a}\Psi_{a}\propto-\frac{\beta^2}{3}{(\rho-3P_{i}+3\zeta\Theta)(\rho+P_{\parallel})},\,\,\,P_{i}=P_{r}=P_{t}.
\end{equation}
Under physically reasonable energy conditions, $(\rho-3P_{i}+3\zeta\Theta) $ determines the nature of the MOTS as $ (\rho+P_{\parallel}) $ is a positive quantity. Therefore, it is interesting that the bulk viscosity plays a crucial role in determining the nature of the MOTS. To check the causal evolution of the MOTS we write

\begin{equation}
\dfrac{\alpha}{\beta}=-\bigg\{\dfrac{2\rho/3-\eta\Sigma}{\rho/3+P_{i}-\zeta\Theta-\eta\Sigma}\bigg\}.
\end{equation}
We can directly conclude from the \textbf{DEC} that the energy density terms dominate over other terms (near the singularity), and therefore, the ratio becomes negative. Therefore, the MOTS becomes locally future ingoing, making the central singularity always covered. This example demonstrates transparently that the Weyl scalar is crucial in the formation of a locally naked singularity.

\subsubsection{Conformally flat spacetimes}

To illustrate the point further in a general viscous setting, we switch off the Weyl scalar in our model to investigate the visibility of the central singularity. Therefore, we will now consider a conformally flat $\mathcal{E}=0$, dissipative, anisotropic (viscous) spacetime.  Consequently, we obtain the following
\begin{equation}
\Psi^{a}\Psi_{a}=-\frac{\beta^2 R}{3(P_{r}-2\eta\Sigma-\zeta\Theta+K-Q)^2}(\rho+P_{r}-2\eta\Sigma-\zeta\Theta-2Q).
\end{equation}
For a Type I matter field and a physically reasonable energy condition, the Ricci scalar controls the nature of the MOTS. Therefore, the bulk viscosity plays a decisive role via the Ricci scalar (the shear viscosity can only influence through the Weyl scalar). The energy density dominates during the end state of the collapse due to \textbf{DEC}. Therefore the ratio

\begin{equation}
\frac{\alpha}{\beta}=\dfrac{-{2\rho}/{3}-\Delta P/{3}+\eta\Sigma+Q}{(2P_{r}+P_{t})/3-\zeta\Theta-\eta\Sigma+ {\rho}/{3}-Q}<0,
\end{equation}
becomes negative, making the MOTS ingoing near the singularity. For a positive (negative) Ricci scalar, the MOTS becomes future ingoing timelike (spacelike) and a zero Ricci scalar would make the MOTS future ingoing null, making the singularity covered. We could also directly argue that, as the MOTS is ingoing, the central singularity is always covered. We can further weaken the Ricci sector by making the fluid isotropic i.e., $ \Pi=0$ or $ \Delta P=3\eta \Sigma$ to show that without the Weyl scalar the singularity will always be covered in a physically reasonable collapse. Therefore, in the isotropic case, the ratio becomes
\begin{equation}
\dfrac{\alpha}{\beta}=\dfrac{-2\rho/3+Q}{P_{t}+\eta\Sigma-\zeta\Theta+\rho/3-Q}<0.
\end{equation}
It is clear that for a reasonable energy condition, the central shell-focusing singularity will remain covered in a general conformally flat spacetime, as the MOTS is always future ingoing in its vicinity.

\section{Concluding remarks}\label{sec8}

In the previous sections we saw how the Ricci scalar and the Weyl scalar compete between each other to determine the nature of the MOTS. The covariant quantity $ \mathcal{D}\equiv R+6\mathcal{E} $ emerges as the decisive quantity that classifies the MOTS and determines the outcome of collapse. When $ \mathcal{D}=0 $ the MOTS is null; for $ \mathcal{D}>0 $ it is timelike; and for 
$ \mathcal{D}<0 $ it is spacelike. The Ricci scalar for the interior of the star is 
\begin{equation}
R=-T=\rho-3p=\rho-P_r-2P_t+3\zeta\Theta.
\end{equation}
Therefore, the bulk viscosity plays a crucial role in determining the causal nature of the MOTS through the Ricci scalar. However, we know the field equations are governed by the full Ricci tensor (along with the Ricci scalar). In order to understand the full extent of the effect of viscosity on the entire collapsing star we obtain the Ricci tensor square $\mathscr{R}\equiv R^{ab}R_{ab}$ as 
\begin{align}
\mathscr{R}=& \rho^2+3p^2+\frac{3\Pi^2}{2}-2Q^2 \nonumber\\
=\, & \rho^2 + P_r^2 + 2P_t^2 - 2Q^2 + 6\eta^2\Sigma^{2}+ 3\zeta^2\Theta^2 -4\eta \Sigma\Delta P \nonumber\\
&  - 2\zeta\Theta(P_r + 2P_t).
\end{align}
We have four extra terms related to the viscosity in the overall curvature (due to matter). Similarly, to understand the free-gravitational field contributions during collapse we need to compute the Weyl tensor square as ${\mathscr{W}}\equiv C^{abcd}C_{abcd}$. In spherical symmetry, $\mathscr{W}$ can be fully expressed in terms of the electric Weyl scalar as $\mathscr{W}=12{\mathcal{E}}^2$, where, using Eq.(\ref{Mass}), we may express the electric Weyl scalar as
\begin{equation}
{\mathcal E}=-\frac{2\mathcal{M}}{{C}^3}+\frac{\rho-\Delta P}{3}+\eta \Sigma.
\end{equation}
The explicit contribution of the shear viscosity in the free-gravitational field energy is evident \cite{R11,R17}. Both shear and bulk viscosity enter explicitly into the master equation that governs the evolution of the Weyl scalar. They influence how the Ricci-Weyl balance tips during collapse. Shear viscosity couples directly to the Weyl curvature, while bulk viscosity modifies the Ricci scalar. In particular, viscosity terms can delay the formation of trapped surfaces, thereby affecting the timing of censorship. Our work shows that viscosity is not a mere dissipative addition but an active player in the curvature dynamics that can alter the final fate of the collapse. However, ultimately everything depends on the delicate balance between $ R$  (or $ -T $) and $6\mathcal{E} $. This transparently demonstrates the competition between the matter sector and free gravity in determining the causal nature of the MOTS. Due to this, on the null MOTS we get an interesting relation among the structure scalars and the complexity factor as well: $ Y_{TF}=Y_{T}+(X_{TF}-X_{T})$ \cite{He2}.

A Ricci dominated MOTS will cover the singularity, while a Weyl (diverging free gravitational field) dominated MOTS will give us a sneak peek into the high curvature region of the central singularity (at least locally) governed by the following ratio
\begin{equation}
\dfrac{\vert\mathcal{E}\vert}{\rho+P_{r}-2\eta\Sigma-\zeta\Theta-2Q}\rightarrow \infty.
\end{equation} 
Therefore, through the master equation Eq.(\ref{master}) and Eq.(\ref{weyl_bound}) we can understand how the Weyl scalar can become unbounded, driven by the shear evolution. This provides us a transparent process of Weyl divergence over the net inertial mass (denominator) term.

We also obtain the causal wave equation of the Gaussian curvature of the 2-shells on the MOTS curve \cite{R33} to understand the movement of the trapped shells on the apparent horizon (outer MOTS) as 
\begin{equation}
\ddot{K}-\hat{\hat{K}}+\mathcal{F}_{MOTS}K=0,
\end{equation}
where 
\begin{equation}
\mathcal{F}_{MOTS}\equiv -2K+\phi(\phi+\Theta+\mathcal{A})+\rho-P_{r}+2\eta\Sigma+\zeta\Theta.
\end{equation}
It is interesting to see that the `restoration factor' $\mathcal{F}_{MOTS}$ of the wave equation depends explicitly on the difference of the energy density and the effective radial pressure (which is proportional to $ \hat{\mathcal{M}}-\dot{\mathcal{M}} $). Therefore, this Ricci component vanishes when we have $ \nabla_{a}\mathcal{M}\nabla^{a}\mathcal{M}=0 $. The above wave equation also demonstrates the effect of shear and bulk viscosity on the propagation and evolution of the Gaussian curvature of shells on the MOTS. However, tangential pressure has no influence on the wave equation explicitly.

To compare the relative strength of the curvature components we can employ the `gravitational epoch function' \cite{arrow1}, which is defined as $ \mathscr{P}={\mathscr{W}}/{\mathscr{R}}. $
Hence, the covariant time evolution of the epoch function can give us an understanding of the relative dynamics of curvature strength (trace versus trace-free part of the Riemann tensor \cite{R16}) during collapse. This epoch function can also be used to test the gravitational arrow of time during collapse (like in cosmology).
\begin{equation}
\dot{\mathscr{P}}=\mathscr{P}\left(2\frac{\dot{\mathcal{E}}}{\mathcal{E}}-\frac{\dot{\mathscr{R}}}{\mathscr{R}}\right),\,\,\,\Rightarrow \dot{\mathscr{P}}>0,\,\, \text{iff} \,\,\,2\frac{\dot{\mathcal{E}}}{\mathcal{E}} > \frac{\dot{\mathscr{R}}}{\mathscr{R}}.
\end{equation}

Therefore, if the Ricci component dominates over the Weyl part during evolution, the gravitational epoch function becomes a decreasing function, making the interior gravitational arrow of time (measured by the epoch function) opposite to the direction of the thermodynamic arrow of time of the exterior outgoing radiation \cite{arrow1,arrow2}. Interestingly, if we have a scenario where (at least momentarily in the vicinity of the central singularity) the Weyl curvature evolution dominates over the Ricci tensor square, then the two arrows of time will align in the same direction. This means that if we allow the weak CCC to persist, then there is a possibility that the gravitational arrow of time perfectly aligns with the exterior radiation arrow of time. However, this is a much stronger restriction and needs more investigation. The existence of a locally naked singularity opens up many interesting features of spacetime and the arrow of time is one of them. This connection, while speculative, hints at a deeper thermodynamic interpretation of the censorship process as it might correspond to a phase of maximal gravitational entropy production, as can be seen from the obtained condition.

The Ricci-Weyl balance framework opens several new lines of inquiry. Our covariant conditions can be directly checked in numerical relativity simulations of viscous, dissipative collapse. Does the critical balance $R+6\mathcal{E}=0$ define a universal class of collapse outcomes across different equations of state?
The divergence of Weyl curvature near the singularity suggests that quantum gravity effects may become important precisely when censorship is threatened and our classical conditions could serve as boundary conditions for semiclassical models. While extreme, the regimes we study might be approached in the collapse of very massive, differentially rotating, or magnetized stars where viscosity and dissipation are significant. As the Weyl scalar controls the gravitational waves and tidal forces, the vicinity of the central singularity (if locally naked) will be so distorted that if the outgoing null rays manage to escape (violating strong CCC) it will completely destroy the asymptotic structure of the spacetime creating all sorts of issues in conventional gravitational theorems. Therefore, from this perspective (given the obtained condition of locally naked singularity) a weak CCC is plausible in our universe.

In summary, the Ricci-Weyl curvature balance provides a unified geometric framework for understanding singularity censorship. It shifts the focus from specific matter models to the underlying curvature competition, offering a fresh perspective. Our results extend earlier covariant censorship analyses to realistic dissipative, viscous collapse and provide clear, testable criteria for future numerical and theoretical investigations.

\section*{Acknowledgements}
SC is thankful to the University of KwaZulu-Natal (UKZN) for post doctoral funding. RG, SDM and GA thanks UKZN for research support.  

\section*{Conflict of interest}
We confirm no conflict of interest exist.
\section*{Data availability statement}
All data that support the findings of this study are included within the article (and any supplementary files).

\end{document}